\documentclass[twocolumn,amsmath,amssymb,10pt,aps]{revtex4}
  
\usepackage[utf8]{inputenc}
\usepackage[T1]{fontenc}

\usepackage{amsmath}
\usepackage{amsfonts}
\usepackage{graphicx}
\usepackage{yfonts}
\usepackage{color}
\usepackage[normalem]{ulem}
\usepackage{amsthm}
\usepackage{bm}
\usepackage{bbm}
\usepackage{mathtools}
\usepackage{array}
\usepackage{placeins}
\usepackage{enumitem}
\usepackage{longtable}

\newcommand{\der}{\mathrm{d}}
\newcommand\bigforall{\mbox{\Large $\mathsurround=1pt\forall$}}

\def\<{\langle}
\def\>{\rangle}

\newcommand{\Tr}{\mathrm{Tr}}
\def\oper{{\mathchoice{\rm 1\mskip-4mu l}{\rm 1\mskip-4mu l}
{\rm 1\mskip-4.5mu l}{\rm 1\mskip-5mu l}}}
\DeclareMathAlphabet\mathbfcal{OMS}{cmsy}{b}{n}

\begin{document}

\title{Geometry of generalized Pauli channels}

\author{Katarzyna Siudzi{\'n}ska}
\affiliation{Institute of Physics, Faculty of Physics, Astronomy and Informatics \\  Nicolaus Copernicus University, Grudzi\k{a}dzka 5/7, 87--100 Toru{\'n}, Poland}

\begin{abstract}
We analyze the geometry of the generalized Pauli channels constructed from the mutually unbiased bases. The Choi-Jamio{\l}kowski isomorphism allows us to express the Hilbert-Schmidt line and volume elements in terms of the eigenvalues of the generalized Pauli maps. After determining appropriate regions of integration, we analytically compute the volume of generalized Pauli channels and their important subclasses. In particular, we obtain the volumes of the generalized Pauli channels that can be generated by a legitimate generator and are entanglement breaking. We also provide the upper bound for the volume of positive, trace-preserving generalized Pauli maps.
\end{abstract}

\flushbottom

\maketitle

\thispagestyle{empty}

\section{Introduction}

Due to their unique properties, both the Pauli channels and the generalized Pauli channels have found many important applications in quantum information theory. Their construction is mainly based on the sets of mutually unbiased bases \cite{Wootters}. The generalization of the Pauli channels, also known as {\it Pauli diagonal channels constant on axes}, was first defined analyzed by Nathanson and Ruskai \cite{Ruskai}. They were interpreted as the depolarizing channels whose degree of depolarization depends on the choice of the axis. Most effort was put into finding the properties of the channels constructed from the maximal number of mutually unbiased bases (MUBs). However, the authors also showed how to define the generalized Pauli channels for non-maximal numbers of MUBs, and these are the channels that are our main consideration.

This paper is a continuation of \cite{Pauli_volume}, where we introduced the geometry on the space of positive, trace-preserving Pauli maps. There, we used the Choi-Jamio{\l}kowski isomorphism \cite{Choi,Jamiolkowski} to obtain the the Hilbert-Schmidt metric on the space of Pauli maps. Using this metric, we calculated the volumes of the entanglement breaking channels, P and CP-divisible channels, as well as the channels obtainable through time-local generators. Other works on the geometry of quantum channels include \cite{Arvind}, where Narang and Arvind obtained the volume of the Pauli channels that can be simulated by a one-qubit mixed state environment. Similar calculations were repeated by Jung et al. for the generalized amplitude damping channels \cite{Jung}. The Choi-Jamio{\l}kowski isomorphism was also used to compute the volume of unital qubit channels with the Lebesque measure \cite{Lovas}. For the Hilbert-Schmidt measure, Szarek et. al. \cite{Szarek2} listed the bounds for the  volume radii of positive, trace-preserving qudit maps and their nested subsets. There also exist first results for continuous variable systems. The geometry of quantum Gaussian channels has been considered with respect to the Hilbert-Schmidt \cite{Gaussians} and Bures-Fisher metric \cite{Monras2010}.

In the following sections, we start with the definition of the generalized Pauli channels in dimension $d$ that are constructed from $N\leq d+1$ mutually unbiased bases. We derive the positivity and complete positivity conditions in terms of the channel eigenvalues. Next, we introduce the associated Choi-Jamio{\l}kowski states to find the Hilbert-Schmidt line and volume elements of the channels. Having obtained the necessary positivity conditions, we give the upper bound on the volume of positive, trace-preserving generalized Pauli maps. It is very hard to establish analytical regions of integration for arbitrary $N$ and $d$, and for this reason we consider three special examples: $N=d+1$, $N=d$, and $N=3$, two of which yield the same results. In each case, we present the integration regions in such a way that it is possible to analytically calculate the volume integrals. This way, we get the relative volumes as functions of dimension $d$ for the generalized Pauli channels, as well as the subset of channels that can are provided with legitimate generator and entanglement breaking channels. Due to the generalized Pauli channels being higher-dimensional, and therefore harder to analyze, than the Pauli channels, some of our results provide bounds on the volumes rather than exact values. In Conclusions, we list the open problems that have arisen during our research.

\section{Generalized Pauli channels}

Consider mixed unitary qubit evolution given by
\begin{equation}\label{Pauli}
\Lambda[\rho]=\sum_{\alpha=0}^3p_\alpha\sigma_\alpha\rho\sigma_\alpha
\end{equation}
where $p_\alpha$ is a probability distribution and $\sigma_\alpha$ denote the Pauli matrices. Observe that $\Lambda$ is unital; i.e., it preserves the maximally mixed state $\rho_\ast=\mathbb{I}_2/2$. An important feature of the Pauli channel is that the eigenbases of the Pauli matrices are mutually unbiased. Recall that two orthonormal bases $\{\psi_\alpha\}$, $\{\varphi_\beta\}$ in $\mathcal{H}\simeq\mathbb{C}^d$ are mutually unbiased if and only if their vectors satisfy
\begin{equation}
|\<\psi_\alpha|\varphi_\beta\>|^2=\frac 1d.
\end{equation}
One can always find between three and $d+1$ mutually unbiased bases in a given dimension $d$ \cite{MUB-2}. The maximal number $N=d+1$ of MUBs is reached for $d$ being a prime number or a prime power \cite{MAX,Wootters}. If $d=d_1d_2$, then there are at least $N=1+\min\{d_1,d_2\}$ mutually unbiased bases \cite{MUB-1}.

Now, let $P_k^{(\alpha)}:=|\psi_k^{(\alpha)}\>\<\psi_k^{(\alpha)}|$ be a rank-1 projector onto the $k$-th vector $|\psi_k^{(\alpha)}\>$ of $\alpha$-th mutually unbiased bases. Then, construct the quantum-classical channels $\Phi_\alpha[\rho]=\sum_{k=0}^{d-1}P_k^{(\alpha)}\rho P_k^{(\alpha)}$.
The generalized Pauli map in any dimension $d$ is constructed from $N\leq d+1$ mutually unbiased bases in the following way \cite{Ruskai},
\begin{equation}
\Lambda=p_{N+1}\oper+\sum_{\alpha=0}^Np_\alpha\Phi_\alpha,
\end{equation}
where $\Phi_0[\rho]:=\frac 1d \mathbb{I}_d\Tr[\rho]$ is the completely depolarizing channel.

For $N<d+1$, the generalized Pauli map $\Lambda$ is a quantum channel if and only if
\begin{equation}\label{CPT}
\left\{
\begin{aligned}
&p_{N+1}+\sum_{\alpha=1}^Np_\alpha+p_0=1,\\
&d^2p_{N+1}+d\sum_{\alpha=1}^Np_\alpha +p_0\geq 0,\\
&dp_\alpha+p_0\geq 0,\\
&p_0\geq 0.
\end{aligned}\right.
\end{equation}
Indeed, $\Phi_0$ and $\Phi_\alpha$ can be alternatively rewritten as
\begin{equation}
\Phi_\alpha[\rho]=\frac 1d \sum_{k=0}^{d-1}U_\alpha^k\rho U_\alpha^k
\end{equation}
and
\begin{equation}\label{Phi0}
\begin{aligned}
\Phi_0[\rho]=\frac{1}{d^2}\Bigg(\rho+\sum_{\alpha=1}^N\sum_{k=1}^{d-1}&U_\alpha^k\rho U_\alpha^{k\dagger}\\&+\sum_{\beta=1}^{d+1-N}\sum_{k=1}^{d-1}A_{\beta,k}\rho A_{\beta,k}^\dagger\Bigg),
\end{aligned}
\end{equation}
respectively.
The set $\{\mathbb{I}_d,U_\alpha^k,A_{\beta,k}\}$ forms an orthogonal operator basis, whereas $U_\alpha^k:=\sum_{j=0}^{d-1}\omega^{jk}P_j^{(\alpha)}$ are the unitary operators constructed from $P_k^{(\alpha)}$ and $\omega:=e^{2\pi i/d}$. Therefore, the complete positivity conditions are equivalent to the positivity of all the coefficients in the operator sum representation, which is exactly (\ref{CPT}). The map $\Lambda$ is mixed unitary if $U_\alpha^k$ are the Weyl operators or their tensor products, as then one can always find unitary operators $A_{\beta,k}$ that complete the orthonormal basis. In this case, the generalized Pauli channel is mixed unitary regardless of the dimension $d$. Using the Weyl operators, one can always construct at least $N=3$ mutually unbiased bases \cite{MUB-2}. If $N=d+1$, then $\oper=\sum_{\alpha=1}^N\Phi_\alpha-d\Phi_0$ is linearly dependent on $\Phi_\alpha$ and $\Phi_0$. Hence, $p_{N+1}$ is set to zero as an excessive degree of freedom and the condition $p_0\geq 0$ is dropped \cite{Ruskai}.

Note that any generalized Pauli channel is unital, as $\Lambda[\mathbb{I}_d]=\mathbb{I}_d$. Moreover,
\begin{equation}
\Phi_\alpha[U_\beta^k]=\delta_{\alpha\beta}U_\beta^k,\qquad 
\Phi_\alpha[A_{\beta,l}]=0,
\end{equation}
and hence the remaining eigenvalue equations read
\begin{equation}
\Lambda[U_\alpha^k]=(p_\alpha+p_{N+1})U_\alpha^k,\qquad\Lambda[A_{\beta,l}]=p_{N+1}A_{\beta,l}.
\end{equation}
The complete positivity conditions in terms of the eigenvalues
\begin{equation}\label{eigenvalues}
\lambda_\alpha=p_{N+1}+p_\alpha,\qquad\lambda_{N+1}=p_{N+1}
\end{equation}
are as follows,
\begin{equation}\label{CP}
\begin{aligned}
-\frac{1}{d-1}&\leq\sum_{\beta=1}^N\lambda_\beta+(d+1-N)\lambda_{N+1}\\&\leq 1+d\min_{\alpha=1,\ldots,N+1}\lambda_\alpha.
\end{aligned}
\end{equation}
Note that for $N=d+1$, one has $\lambda_{N+1}=0$, which recovers the generalized Fujiwara-Algoet conditions \cite{Ruskai,mub_final}. Moreover, conditions for $N=d$ and $N=d+1$ coincide.

\section{Line and volume elements}

Let us analyze the geometry of the generalized Pauli channels $\Lambda$ by considering the properties of the associated Choi-Jamio{\l}kowski quantum states \cite{Choi,Jamiolkowski}
\begin{equation}
\rho_\Lambda=\frac 1d \sum_{k,l=0}^{d-1}|k\>\<l|\otimes\Lambda[|k\>\<l|].
\end{equation}
In general, it is very hard to explicitly write the state $\rho_\Lambda$ corresponding to a given $\Lambda$, as the choice of the mutually unbiased bases is arbitrary. However, we can still introduce the metric via the Hilbert-Schmidt line element $\der s^2=\Tr(\der\rho_\Lambda^2)$ \cite{Sommers2}. In terms of the channel eigenvalues, it reads
\begin{equation}
\der s^2=\frac{d-1}{d^2}\Big(\sum_{\alpha=1}^N\der\lambda_\alpha^2
+(d+1-N)\der\lambda_{N+1}^2\Big).
\end{equation}
Now, the metric tensor has the diagonal form $g=\frac{d-1}{d^2}\mathrm{diag}(1,\ldots,1,d+1-N)$ with $N$ ones. Hence, the associated volume element $\der V=\sqrt{\det g}\prod_{\alpha=1}^{N+1}\der\lambda_k$ is given by
\begin{equation}
\der V=\sqrt{d+1-N}\left(\frac{\sqrt{d-1}}{d}\right)^{N+1}\prod_{\alpha=1}^{N+1}\der\lambda_\alpha.
\end{equation}
Notably, for $N=d+1$, $\der\lambda_{N+1}^2$ vanishes, and therefore one has
\begin{equation}
\begin{aligned}
\der s^2&=\frac{d-1}{d^2}\sum_{\alpha=1}^{d+1}\der\lambda_\alpha^2,\\
\der V&=
\left(\frac{\sqrt{d-1}}{d}\right)^{d+1}\prod_{\alpha=1}^{d+1}\der\lambda_\alpha.\label{dV}
\end{aligned}
\end{equation}
For $d=2$, the above formulas reproduce the line and volume elements for the Pauli channels that have been recently obtained in \cite{Pauli_volume}.

\section{Relative volumes of quantum channels}

In this section, we would like to analytically derive the volumes of the generalized Pauli channels and their important subclasses. However, it is a non-trivial task to find the regions of integration for arbitrary $N$ and $d$. Therefore, for the most part, we limit our consideration to two extreme cases: $N=d+1$ and $N=3$.

\subsection{Generalized Pauli positive maps}

Before we calculate the volumes of generalized Pauli channels, let us find the normalization factor that makes our results more clear to read. Consider a trace-preserving generalized Pauli map $\Lambda$ that, in general, is not completely positive. From definition, $\Lambda$ is positive if and only if
\begin{equation}\label{defGP}
\bigforall_{P,Q}\quad\Tr(\Lambda[Q]P)\geq 0,
\end{equation}
where $P$ and $Q$ are rank-1 projectors. Unfortunately, this inequality is very hard to check. Therefore, we analyze the necessary conditions
\begin{equation}\label{def2GP}
\min_{P,Q\in\{P_k^{(\alpha)},Q_l^{(\beta)}\}}\Tr(\Lambda[Q]P)\geq 0
\end{equation}
with the minimum taken over
\begin{equation}
P_k^{(\alpha)}=\frac 1d \left[\mathbb{I}_d+\sum_{m=1}^{d-1}\omega^{-mk}U_\alpha^m\right]
\end{equation}
and
\begin{equation}
Q_l^{(\beta)}=\frac 1d \left[\mathbb{I}_d+\sum_{n=1}^{d-1}q_n^{(\beta,l)}A_{\beta,n}\right].
\end{equation}
Using the method presented in \cite{norms}, it is straightforward to show that
\begin{equation}\label{CCGP}
\begin{aligned}
\min_{P,Q\in\{P_k^{(\alpha)},Q_l^{(\beta)}\}}&\Tr(\Lambda[Q]P)\\&=\frac 1d \Big[1+\min\{-\lambda_{\max},(d-1)\lambda_{\min}\}\Big],
\end{aligned}
\end{equation}
where $\lambda_{\max}=\max_{\alpha}\lambda_\alpha$ and $\lambda_{\min}=\min_{\alpha}\lambda_\alpha$ for $\alpha=1,\ldots,N$. Note that for $N=d+1$, the above formula is the maximal output $\infty$-norm \cite{norms}.
Now, the necessary condition for positivity of trace-preserving generalized Pauli maps is
\begin{equation}\label{CCGP2}
-\frac{1}{d-1}\leq\lambda_\alpha\leq 1.
\end{equation}
Unfortunately, numerical calculations indicate that such conditions may lead to maps that are not positive. Hence, condition (\ref{def2GP}) is not sufficient. However, we can at least calculate the upper estimate on the volume of positive, trace-preserving generalized Pauli maps. Explicitly, it reads
\begin{equation}\label{VP_dN}
\begin{aligned}
V_P(d,N)&=\sqrt{d+1-N}\left(\frac{\sqrt{d-1}}{d}\right)^{N+1}\prod_{\alpha=1}^{N+1}
\int_{-\frac{1}{d-1}}^1\der\lambda_\alpha\\
&=\sqrt{\frac{d+1-N}{(d-1)^{N+1}}}
\end{aligned}
\end{equation}
for $N=1,\ldots,d$ and
\begin{equation}\label{VP_d}
\begin{aligned}
V_P(d)&=\left(\frac{\sqrt{d-1}}{d}\right)^{d+1}\prod_{\alpha=1}^{d+1}
\int_{-\frac{1}{d-1}}^1\der\lambda_\alpha\\
&=\frac{1}{\sqrt{(d-1)^{d+1}}}
\end{aligned}
\end{equation}
for $N=d+1$. Note that for $N=d$, one has $V_P(d,d)=V_P(d)$. In the upcoming subsection, $V_P(d)$ is used for normalization purposes.

\subsection{$N=d+1$ and $N=d$ mutually unbiased bases}

From now on, let us assume, without a loss of generality, that $\lambda_1\leq\lambda_2\leq\ldots\leq\lambda_{d+1}$. Note that there are $(d+1)!$ possible index permutations for a given dimension $d$, so the total volume for arbitrarily ordered $\lambda_\alpha$ is $(d+1)!$ times the volume for non-decreasingly ordered eigenvalues (for details, see Appendix B). This simple trick significantly simplifies the regions of integration. The results of this subsection are calculated for $N=d+1$, but they also apply to the case with $N=d$.

The volume $V_{CP}(d)$ of the generalized Pauli channels is calculated by integrating the volume element $\der V$ over the complete positivity region (see Appendix \ref{A}). Due to the nature of lower and upper bounds of $\lambda_\alpha$, we are unable to derive the final formula for $V_{CP}(d)$ directly. Instead, we find the analytical volume ratios $V_{CP}/V_P$ for small values of $d$ (for more details, see Appendix B),
\begin{equation*}
\begin{aligned}
\frac{V_{CP}(2)}{V_P(2)}=\frac 13,\qquad
&\frac{V_{CP}(4)}{V_P(4)}=\frac{1}{30},\\
\frac{V_{CP}(3)}{V_P(3)}=\frac 18,\qquad&\frac{V_{CP}(5)}{V_P(5)}=\frac{1}{144},
\end{aligned}
\end{equation*}
and conjecture that the general formula reads
\begin{equation}
\frac{V_{CP}(d)}{V_P(d)}=\frac{d}{(d+1)!}.
\end{equation}

Quantum channels are used to provide the evolution of open quantum systems. Instead of a single channel $\Lambda$, one takes a time-parametrized family $\{\Lambda(t)|t\geq 0,\Lambda(0)=\oper\}$ that is called the {\it dynamical map}. The simplest example of a dynamical map is the Markovian semigroup $\Lambda(t)=e^{t\mathcal{L}}$. It is the solution of the master equation
\begin{equation}\label{MS}
\dot{\Lambda}(t)=\mathcal{L}\Lambda(t),
\end{equation}
where $\mathcal{L}$ is the Gorini-Kossakowski-Sudarshan-Lindblad (GKSL) generator \cite{GKS,L}
\begin{equation}\label{L}
\mathcal{L}=\sum_{\alpha=0}^N\gamma_\alpha(\Phi_\alpha-\oper)
\end{equation}
with positive decoherence rates $\gamma_\alpha\geq 0$. If $N=d+1$, then $\gamma_0=0$ \cite{mub_final}. One way of including non-Markovian effects is to introduce the time-local generator $\mathcal{L}(t)$, which is given via formula (\ref{L}) but with time-dependent $\gamma_\alpha(t)\ngeq 0$. Observe that the generalized Pauli channel $\Lambda$ belongs to a family of generalized Pauli dynamical maps $\Lambda(t)$ generated by $\mathcal{L}(t)$ if and only if $\Lambda=\Lambda(t_\ast)$ for a fixed $t_\ast\geq 0$. Equivalently, one has $\lambda_\alpha\geq 0$, which provides the integration region
\begin{equation}
\left\{
\begin{aligned}
&\lambda_{k-1}\leq\lambda_k\leq\frac{1}{d+2-k}\left(1+d\lambda_1
-\sum_{\beta=1}^{k-1}\lambda_\beta\right),\\
&\qquad\qquad\qquad\qquad\qquad\qquad k=2,\ldots,d+1,\\
&0\leq\lambda_1\leq 1,
\end{aligned}
\right.
\end{equation}
when paired with the complete positivity condition from Appendix A. Let us call the channels with $\lambda_\alpha\geq 0$ {\it achievable with the time-local generators} and denote their volume by $V_G(d)$. Again, we calculate the analytical volumes $V_G(d)$ for small values of $d$,
\begin{equation*}
\begin{aligned}
\frac{V_{G}(2)}{V_{CP}(2)}=\frac{3}{16},\qquad
&\frac{V_{G}(4)}{V_{CP}(4)}=\frac{1215}{4096},\\
\frac{V_{G}(3)}{V_{CP}(3)}=\frac{64}{243},\qquad
&\frac{V_{G}(5)}{V_{CP}(5)}=\frac{24576}{78125},
\end{aligned}
\end{equation*}
and conjecture that the general formula reads
\begin{equation}\label{G}
\frac{V_{G}(d)}{V_{CP}(d)}=\frac{d^2-1}{d^2}\left(\frac{d-1}{d}\right)^{d}.
\end{equation}
Surprisingly, this is the only relative volume we consider that does not go to zero with the increase of $d$. Indeed, if we take the limit $d\to\infty$, then eq. (\ref{G}) monotonically approaches $e$ from below.

An interesting subclass of the generalized Pauli channels, with many applications in quantum communication and information processing, consists in the channels that are entanglement breaking (EB). A quantum channel $\Lambda$ is entanglement breaking if and only if its Choi-Jamio{\l}kowski state $\rho_\Lambda$ is separable. If $\lambda_\alpha\geq 0$, then $\sum_{\alpha=1}^{d+1}\lambda_\alpha\leq 1$ is the necessary and sufficient condition for breaking the entanglement \cite{Ruskai}. Therefore, the generalized Pauli channel achievable with the time-local generators that are also EB corresponds to the region
\begin{equation}
\left\{
\begin{aligned}
&\lambda_{k-1}\leq\lambda_k\leq\frac{1}{d+2-k}\left(1
-\sum_{\beta=1}^{k-1}\lambda_\beta\right),\\
&\qquad\qquad\qquad\qquad\qquad\qquad k=2,\ldots,d+1,\\
&0\leq\lambda_1\leq\frac{1}{d+1}.
\end{aligned}
\right.
\end{equation}
For small dimensions $d$, we find relatively simple analytical formulas for the volumes $V_{EB}(d)$ of the entanglement breaking channels achievable with the time-local generators,
\begin{equation*}
\begin{aligned}
\frac{V_{EB}(2)}{V_{G}(2)}=\frac{1}{3},\qquad
&\frac{V_{EB}(4)}{V_{G}(4)}=\frac{1}{5},\\
\frac{V_{EB}(3)}{V_{G}(3)}=\frac{1}{4},\qquad
&\frac{V_{EB}(5)}{V_{G}(5)}=\frac{1}{6},
\end{aligned}
\end{equation*}
from which we conjecture that
\begin{equation}
\frac{V_{EB}(d)}{V_{G}(d)}=\frac{1}{d+1}.
\end{equation}

\begin{figure}[ht!]
  \includegraphics[width=0.4\textwidth]{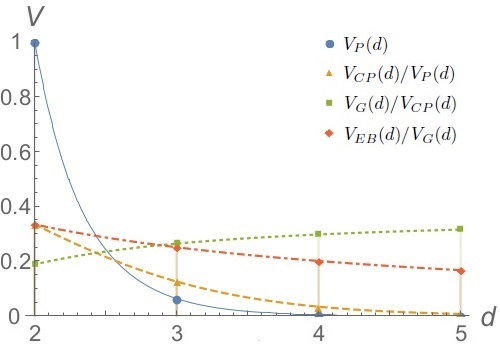}
\caption{A graphical representation of the volumes for selected classes of the generalized Pauli trace-preserving maps constructed from $d$ or $d+1$ mutually unbiased bases.}
\label{max}
\end{figure}

The results of this subsection are presented in Fig. \ref{max}. The curves have no physical meaning and are plotted only to show the behavior of the discretely-valued relative volumes. Note that the only function that does not monotonically decrease with the increase of the dimension $d$ is $V_{G}(d)/V_{CP}(d)$, which approaches $1/e$ for $d\to\infty$. This means that the bigger the Hilbert space, the more there are generalized Pauli channels with positive eigenvalues. For $d=2$, the results coincide with \cite{Pauli_volume}. Interestingly, for $d>2$, the volume of positive, trace-preserving maps is no longer normalized.

\subsection{$N=3$ mutually unbiased bases}

Now, let us analyze the volumes of the generalized Pauli channels constructed from three mutually unbiased bases. This is another interesting choice, as one can construct three MUBs in any dimension. Once again, in order to simplify the integrals, we assume that $\lambda_1\leq\lambda_2\leq\lambda_3$. There are $3!=6$ possible index permutations for $\lambda_\alpha$, $\alpha=1,2,3$. However, due to $N<d+1$, we have one more eigenvalue, $\lambda_4$, and there are four possible ways that $\lambda_4$ can be placed in the above sequence of inequalities. For each possible placement of $\lambda_4$, we derive the integration regions separately (see Appendix \ref{B}). Therefore, the final result is the sum of four integrals multiplied by six.

As the number of eigenvalues is set, we are able to directly derive the formulas for the relative volumes of the channels considered in the previous subsection. These are
\begin{equation}
\begin{aligned}
&V_P(d)=\frac{\sqrt{d-2}}{(d-1)^2},\\
&\frac{V_{CP}(d)}{V_P(d)}=\frac{d}{24(d-2)},\\
&\frac{V_{G}(d)}{V_{CP}(d)}=\frac{(d^2-1)(d-1)^3}{d^5},\\
&\frac{V_{EB}(d)}{V_{G}(d)}=\frac{1}{d+1},
\end{aligned}
\end{equation}
where we used the notation $V(d,3)\equiv V(d)$ for the sake of simplicity. The only difference is that $V_{EB}(d)$ is the volume of channels achievable with the time-local generator that satisfy condition \cite{Ruskai}
\begin{equation}\label{EB3}
\sum_{\alpha=1}^{N}\lambda_\alpha+(d+1-N)\lambda_{N+1}\leq 1
\end{equation}
for breaking the entanglement, which is necessary but may not be sufficient. Hence, for $d=3$, $V_{EB}(d)$ gives just the upper bound for the volume of enranglement breaking channels with positive eigenvalues. Despite this fact, $V_{EB}(d)/V_G(d)$ for $N=3$ is given by the same formula as in the case of $N=d,d+1$, which could indicate that condition (\ref{EB3}) is sufficient, after all.

\begin{figure}[ht!]
  \includegraphics[width=0.4\textwidth]{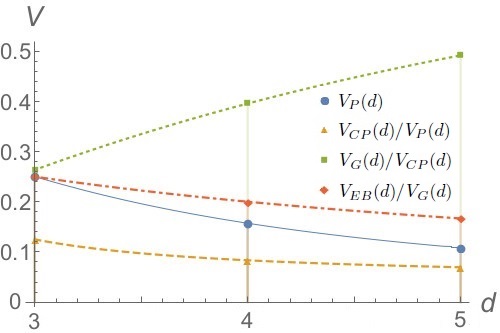}
\caption{
A graphical representation of the volumes for selected classes of the generalized Pauli trace-preserving maps constructed from three mutually unbiased bases.}
\label{min}
\end{figure}

In Fig. \ref{min}, we plot the relative volumes for small dimensions $d$. One can notice several similarities with Fig. \ref{max}. Again, all the functions decrease monotonically with the increase of the dimension $d$, except for $V_{G}(d)/V_{CP}(d)$, which this time approaches identity for $d\to\infty$. For $d=3$, the results agree with the calculations for the maximal number of MUBs from the previous subsection.

\section{Conclusions}

In this paper, we establish the geometry of the generalized Pauli channels constructed from $N$ mutually unbiased bases, where $N$ ranges up to $d+1$. Using the Choi-Jamio{\l}kowski isomorphism between quantum channels and quantum states, we find the Hilbert-Schmidt line and volume elements in the state space. In more details, we consider two special examples: $N=d+1$ and $N=3$. In both cases, we express the integration regions for the channels, channels achievable with the time-local generators and their entanglement breaking subclass in a way that allows us to calculate the volume integrals analytically. Finally, we obtain the relative channel volumes as functions of the dimension $d$. Some interesting behavior can be observed, like the percentage of channels with positive eigenvalues approaching a constant, non-zero value with the increase of $d$. Our results can be interpreted as the probability that a randomly selected generalized Pauli channel has the desired properties.

One of the open problems is the characterization of the integration regions for the generalized Pauli channels constructed from any number of mutually unbiased bases. It would be interesting to obtain the channel volumes with both $d$ and $N$-dependence, even for $\lambda_{N+1}=0$. Tighter conditions for the positive maps and entanglement breaking channels would make it possible to find the exact volumes instead of their upper bounds. Another task is to compare our results with the volumes calculated using other metrics, like the Fisher-Rao metric \cite{Fisher}.

\section*{Acknowledgements}

This paper was supported by the Polish National Science Centre project No. 2018/31/N/ST2/00250.

\bibliography{C:/Users/cynda/OneDrive/Fizyka/bibliography}
\bibliographystyle{C:/Users/cynda/OneDrive/Fizyka/beztytulow2}

\onecolumngrid
\appendix

\section{Regions of integration for $N=d+1$}\label{A}

The complete positivity region for trace-preserving generalized Pauli maps is obtained from the generalized Fujiwara-Algoet condition
\begin{equation}
-\frac{1}{d-1}\leq\sum_{\alpha=1}^{d+1}\lambda_\alpha\leq 1+d\min_\alpha\lambda_\alpha.
\end{equation}
For non-decreasingly ordered channel eigenvalues, this region consists in $\lambda_\alpha$ that satisfy one of the following systems of inequalities:
\begin{equation}
\left\{
\begin{aligned}
&-\frac{1}{d-1}-\sum_{\beta=1}^d\lambda_\beta\leq\lambda_{d+1}\leq 1+d\lambda_1-\sum_{\beta=1}^d\lambda_\beta,\\
&\lambda_{k-1}\leq\lambda_k\leq-\frac{1}{d+2-k}\left(\frac{1}{d-1}
+\sum_{\beta=1}^{k-1}\lambda_\beta\right)\quad\bigforall_{2\leq k\leq d},\\
&-\frac{1}{d-1}\leq\lambda_1\leq-\frac{1}{d^2-1},
\end{aligned}
\right.
\end{equation}
or
\begin{equation}
\left\{
\begin{aligned}
&\lambda_d\leq\lambda_{d+1}\leq 1+d\lambda_1-\sum_{\beta=1}^d\lambda_\beta,\\
&\lambda_{k-1}\leq\lambda_k\leq\frac{1}{d+2-k}\left(1+d\lambda_1
-\sum_{\beta=1}^{k-1}\lambda_\beta\right)\quad\bigforall_{k>M},\\
&-\frac{1}{d+2-M}\left(\frac{1}{d-1}+\sum_{\beta=1}^{M-1}\lambda_\beta\right)
\leq\lambda_M\leq\frac{1}{d+2-M}\left(1+d\lambda_1
-\sum_{\beta=1}^{M-1}\lambda_\beta\right),\\
&\lambda_{k-1}\leq\lambda_k\leq-\frac{1}{d+2-k}\left(\frac{1}{d-1}
+\sum_{\beta=1}^{k-1}\lambda_\beta\right)\quad\bigforall_{k<M},\\
&-\frac{1}{d-1}\leq\lambda_1\leq-\frac{1}{d^2-1},
\end{aligned}
\right.
\end{equation}
for a fixed $M$ such that $2\leq M\leq d$, or
\begin{equation}
\left\{
\begin{aligned}
&\lambda_{k-1}\leq\lambda_k\leq\frac{1}{d+2-k}\left(1+d\lambda_1
-\sum_{\beta=1}^{k-1}\lambda_\beta\right)\quad\bigforall_{2\leq k\leq d+1},\\
&-\frac{1}{d^2-1}\leq\lambda_1\leq 1.
\end{aligned}
\right.
\end{equation}

\section{Analytical integration of the volume}

Let us observe how the analytical volume integrals are calculated by analyzing the example of the Pauli channels, which correspond to the choice $d=2$ and $N=d+1$. Their volume is obtained by integrating the volume element $\der V$ from eq. (\ref{dV}) over the complete positivity region $\mathcal{CP}(2)$,
\begin{equation}
V_{CP}(2)=\int_{\mathcal{CP}(2)}\der V=\frac 18 \int_{\mathcal{CP}(2)}\der\lambda_1
\der\lambda_2\der\lambda_3.
\end{equation}
Now, assume that the eigenvalues are non-decreasingly ordered, so that $\lambda_1\leq\lambda_2\leq\lambda_3$. According to Appendix A, the region of integration $\mathcal{CP}^\prime(2)$ for ordered $\lambda_\alpha$'s is provided by the following systems of inequalities,
\begin{equation}
\left\{
\begin{aligned}
&-1-\lambda_1-\lambda_2\leq\lambda_3\leq 1+\lambda_1-\lambda_2,\\
&\lambda_1\leq\lambda_2\leq-\frac 12 \left(1+\lambda_1\right),\\
&-1\leq\lambda_1\leq-\frac 13,
\end{aligned}
\right.
\quad
\left\{
\begin{aligned}
&\lambda_2\leq\lambda_3\leq 1+\lambda_1-\lambda_2,\\
&-\frac 12\left(1+\lambda_1\right)
\leq\lambda_2\leq\frac 12 \left(1+\lambda_1\right),\\
&-1\leq\lambda_1\leq-\frac 13,
\end{aligned}
\right.
\quad
\left\{
\begin{aligned}
&\lambda_2\leq\lambda_3\leq 1+\lambda_1-\lambda_2,\\
&\lambda_1\leq\lambda_2\leq\frac 12 \left(1+\lambda_1\right),\\
&-\frac 13 \leq\lambda_1\leq 1.
\end{aligned}
\right.
\end{equation}
Therefore, the corresponding volume reads
\begin{equation}
\begin{split}
V_{CP}^\prime(2)=&\int_{\mathcal{CP}^\prime(2)}\der V=\frac 18\Bigg(\int_{-1}^{-\frac 13}\der\lambda_1\int_{\lambda_1}^{-\frac 12 (1+\lambda_1)}\der\lambda_2\int_{-1-\lambda_1-\lambda_2}^{1+\lambda_1-\lambda_2}\der\lambda_3
+\int_{-1}^{-\frac 13}\der\lambda_1\int_{-\frac 12 (1+\lambda_1)}^{\frac 12 (1+\lambda_1)}\der\lambda_2\int_{\lambda_2}^{1+\lambda_1-\lambda_2}\der\lambda_3\\&
+\int_{-\frac 13}^{1}\der\lambda_1\int_{\lambda_1}^{\frac 12 (1+\lambda_1)}\der\lambda_2\int_{\lambda_2}^{1+\lambda_1-\lambda_2}\der\lambda_3\Bigg)
=\frac 18 \cdot\frac 49=\frac{1}{18}
\end{split}
\end{equation}
where the more detailed calculations are omitted due to the integrals being elementary. To obtain the total volume of the Pauli channels $V_{CP}(2)$, we make another observation. Both the volume element $\der V$ and the Fujiwara-Algoet conditions \cite{Fujiwara}
\begin{equation}
-1\leq\lambda_1+\lambda_2+\lambda_3\leq 1+2\min_\alpha\lambda_\alpha
\end{equation}
are symmetric with respect to index permutations. Therefore, by permuting the indices in the ordering $\lambda_1\leq\lambda_2\leq\lambda_3$, we can cover all the possible orders that constitute to the general case of non-ordered $\lambda_\alpha$'s. There are a total of six possible ways to order the channel eigenvalues, and hence we finally get
\begin{equation}
V_{CP}(2)=6V_{CP}^\prime(2)=\frac 13.
\end{equation}

Analogical calculations are performed to derive all the other volumes presented in this paper.
For higher dimensions $d$, the volume $V_{CP}^\prime(d)=\int_{\mathcal{CP}^\prime(d)}\der V$ for non-decreasingly ordered eigenvalues is a sum of $d+1$ polynomial integrals. The region of integration $\mathcal{CP}^\prime(d)$ is again given in Appendix A. Now, there are $(d+1)!$ possible index permutations for ordered $\lambda_\alpha$, so the total volume
$V_{CP}(d)=(d+1)!V_{CP}^\prime(d)$.
The difficulty in obtaining the general formula for $V_{CP}(d)$ lies not in the complexity of integrals but in the fact that this formula is a sum of $d+1$-tuple integrals whose limits depend on $d$. However, for fixed $d$, the derivations, even if lengthy, are straightforward.

\section{Regions of integration for $N=3$}\label{B}

For trace-preserving generalized Pauli maps constructed from three mutually unbiased bases, the complete positivity region is determined by the generalized Fujiwara-Algoet condition
\begin{equation}\label{FA3}
-\frac{1}{d-1}\leq\lambda_1+\lambda_2+\lambda_3+(d-2)\lambda_4\leq 1+d\min\{\lambda_1,\lambda_2,\lambda_3,\lambda_4\}.
\end{equation}
We assume that $\lambda_1\leq\lambda_2\leq\lambda_3$ and
\begin{equation}
\{\lambda_1,\lambda_2,\lambda_3,\lambda_4\}=\{\lambda_{\min},\lambda_{\rm mid1},\lambda_{\rm mid2},\lambda_{\max}\},
\end{equation}
where $\lambda_{\min}\leq\lambda_{\rm mid1}\leq\lambda_{\rm mid2}\leq\lambda_{\max}$. Now, inequality (\ref{FA3}) holds if and only if the eigenvalues $\lambda_\alpha$ satisfy one of the following systems of inequalities:
\begin{equation}
\left\{
\begin{aligned}
&\lambda_{\rm mid2}\leq\lambda_{\max}\leq\frac{1}{1+(d-3)\delta_{\max,4}}\Big[1+(d-1)\lambda_{\min}
-\lambda_{\rm mid1}-\lambda_{\rm mid2}-(d-3)(\delta_{\rm mid1,4}+\delta_{\rm mid2,4})\lambda_4\Big],\\
&\lambda_{\rm mid1}\leq\lambda_{\rm mid2}\leq\frac{1+(d-1)\lambda_{\min}
-\lambda_{\rm mid1}-(d-3)(\delta_{\rm mid1,4}+\delta_{\min,4})\lambda_4}
{d-1-(d-3)(\delta_{\rm mid1,4}+\delta_{\min,4})},\\
&\lambda_{\min}\leq\lambda_{\rm mid1}\leq\frac{1+(d-1)\lambda_{\min}
-(d-3)\delta_{\min,4}\lambda_4}{d-(d-3)\delta_{\min,4}},\\
&-\frac{1}{d^2-1}\leq\lambda_{\min}\leq 1
\end{aligned}
\right.
\end{equation}
or
\begin{equation}
\left\{
\begin{aligned}
&-\frac{1}{1+(d-3)\delta_{\max,4}}\Big[\frac{1}{d-1}+\lambda_{\min}
+\lambda_{\rm mid1}+\lambda_{\rm mid2}+(d-3)(1-\delta_{\max,4})\lambda_4\Big]
\leq\lambda_{\max}\\&\qquad\leq\frac{1}{1+(d-3)\delta_{\max,4}}\Big[1+(d-1)\lambda_{\min}
-\lambda_{\rm mid1}-\lambda_{\rm mid2}-(d-3)(\delta_{\rm mid1,4}+\delta_{\rm mid2,4})\lambda_4\Big],\\
&\lambda_{\rm mid1}\leq\lambda_{\rm mid2}\leq-\frac{\frac{1}{d-1}+\lambda_{\min}
+\lambda_{\rm mid1}+(d-3)(\delta_{\rm mid1,4}+\delta_{\min,4})\lambda_4}
{d-1-(d-3)(\delta_{\rm mid1,4}+\delta_{\min,4})},\\
&\lambda_{\min}\leq\lambda_{\rm mid1}\leq-\frac{\frac{1}{d-1}+\lambda_{\min}
+(d-3)\delta_{\min,4}\lambda_4}{d-(d-3)\delta_{\min,4}},\\
&-\frac{1}{d-1}\leq\lambda_{\min}\leq-\frac{1}{d^2-1}
\end{aligned}
\right.
\end{equation}
or
\begin{equation}
\left\{
\begin{aligned}
&\lambda_{\rm mid2}\leq
\lambda_{\max}\leq\frac{1}{1+(d-3)\delta_{\max,4}}\Big[1+(d-1)\lambda_{\min}
-\lambda_{\rm mid1}-\lambda_{\rm mid2}-(d-3)(\delta_{\rm mid1,4}+\delta_{\rm mid2,4})\lambda_4\Big],\\
&-\frac{\frac{1}{d-1}+\lambda_{\min}+\lambda_{\rm mid1}+(d-3)(\delta_{\rm mid1,4}+\delta_{\min,4})\lambda_4}
{d-1-(d-3)(\delta_{\rm mid1,4}+\delta_{\min,4})}
\leq\lambda_{\rm mid2}\leq\frac{1+(d-1)\lambda_{\min}
-\lambda_{\rm mid1}-(d-3)(\delta_{\rm mid1,4}+\delta_{\min,4})\lambda_4}
{d-1-(d-3)(\delta_{\rm mid1,4}+\delta_{\min,4})},\\
&\lambda_{\min}\leq\lambda_{\rm mid1}\leq
-\frac{\frac{1}{d-1}+\lambda_{\min}+(d-3)\delta_{\min,4}\lambda_4}{d-(d-3)\delta_{\min,4}},\\
&-\frac{1}{d-1}\leq\lambda_{\min}\leq-\frac{1}{d^2-1}
\end{aligned}
\right.
\end{equation}
or
\begin{equation}
\left\{
\begin{aligned}
&\lambda_{\rm mid2}\leq\lambda_{\max}\leq
\frac{1}{1+(d-3)\delta_{\max,4}}\Big[1+(d-1)\lambda_{\min}
-\lambda_{\rm mid1}-\lambda_{\rm mid2}-(d-3)(\delta_{\rm mid1,4}+\delta_{\rm mid2,4})\lambda_4\Big],\\
&\lambda_{\rm mid1}\leq\lambda_{\rm mid2}\leq\frac{1+(d-1)\lambda_{\min}
-\lambda_{\rm mid1}-(d-3)(\delta_{\rm mid1,4}+\delta_{\min,4})\lambda_4}
{d-1-(d-3)(\delta_{\rm mid1,4}+\delta_{\min,4})},\\
&-\frac{\frac{1}{d-1}+\lambda_{\min}+(d-3)\delta_{\min,4}\lambda_4}{d-(d-3)\delta_{\min,4}}\leq
\lambda_{\rm mid1}\leq\frac{1+(d-1)\lambda_{\min}
-(d-3)\delta_{\min,4}\lambda_4}{d-(d-3)\delta_{\min,4}},\\
&-\frac{1}{d-1}\leq\lambda_{\min}\leq-\frac{1}{d^2-1}.
\end{aligned}
\right.
\end{equation}
By $\delta_{\alpha,k}$, we understand the Kronecker delta, where $\alpha=1,2,3,4$ and $k=\min,{\rm mid1},{\rm mid2},\max$. For example, $\delta_{\min,1}$ does not vanish if and only if $\lambda_1=\lambda_{\min}$.
Now, if one has $\lambda_\alpha\geq 0$, then the region of integration for the channels achievable with the time-local generators is determined by
\begin{equation}
\left\{
\begin{aligned}
&\lambda_{\rm mid2}\leq\lambda_{\max}\leq\frac{1}{1+(d-3)\delta_{\max,4}}\Big[1+(d-1)\lambda_{\min}
-\lambda_{\rm mid1}-\lambda_{\rm mid2}-(d-3)(\delta_{\rm mid1,4}+\delta_{\rm mid2,4})\lambda_4\Big],\\
&\lambda_{\rm mid1}\leq\lambda_{\rm mid2}\leq\frac{1+(d-1)\lambda_{\min}
-\lambda_{\rm mid1}-(d-3)(\delta_{\rm mid1,4}+\delta_{\min,4})\lambda_4}
{d-1-(d-3)(\delta_{\rm mid1,4}+\delta_{\min,4})},\\
&\lambda_{\min}\leq\lambda_{\rm mid1}\leq\frac{1+(d-1)\lambda_{\min}
-(d-3)\delta_{\min,4}\lambda_4}{d-(d-3)\delta_{\min,4}},\\
&0\leq\lambda_{\min}\leq 1.
\end{aligned}
\right.
\end{equation}
Finally, if the non-decreasingly ordered eigenvalues of the generalized Pauli channel are non-negative and satisfy condition (\ref{EB3}), then
\begin{equation}
\left\{
\begin{aligned}
&\lambda_{\rm mid2}\leq\lambda_{\max}\leq\frac{1}{1-(d-3)\delta_{\max,4}\lambda_4}\Big[1
-\lambda_{\min}-\lambda_{\rm mid1}-\lambda_{\rm mid2}-(d-3)(1-\delta_{\max,4})\Big],\\
&\lambda_{\rm mid1}\leq\lambda_{\rm mid2}\leq\frac{1-\lambda_{\min}-\lambda_{\rm mid1}-(d-3)(\delta_{\min,4}+\delta_{\rm mid1,4})\lambda_4}
{d-1-(d-3)(\delta_{\min,4}+\delta_{\rm mid1,4})},\\
&\lambda_{\min}\leq\lambda_{\rm mid1}\leq\frac{1-\lambda_{\min}
-(d-3)\delta_{\min,4}\lambda_4}{d-(d-3)\delta_{\min,4}},\\
&0\leq\lambda_{\min}\leq\frac{1}{d+1}.
\end{aligned}
\right.
\end{equation}
Note that there are some similarities between there conditions and the conditions in Appendix \ref{A} -- for example, if $d=3$, the integration regions for $N=3$ and $N=d,d+1$ coincide.

\end{document}